\documentclass[aps,prl,floatfix,superscriptaddress,reprint]{revtex4-2}

\usepackage{graphicx}
\usepackage{amsfonts}
\usepackage{amssymb}
\usepackage{amsmath}
\usepackage{hyperref}

\begin{document}

\title{Finite-size Topology}

\author{Ashley M. Cook}
\affiliation{Max Planck Institute for Chemical Physics of Solids, N{\"o}thnitzer Stra{\ss}e 40, 01187 Dresden, Germany}
\affiliation{Max Planck Institute for the Physics of Complex Systems, N{\"o}thnitzer Stra{\ss}e 38, 01187 Dresden, Germany}
\author{Anne E. B. Nielsen}
\affiliation{Department of Physics and Astronomy, Aarhus University, 8000 Aarhus C, Denmark}

\begin{abstract}
We show that topological characterization and classification in $D$-dimensional systems, which are thermodynamically large in only $D-\delta$ dimensions and finite in size in $\delta$ dimensions, is fundamentally different from that of systems thermodynamically large in all $D$-dimensions: as $(D-\delta)$- dimensional topological boundary states permeate into a system's $D$ dimensional bulk with decreasing system size, they hybridize to create novel topological phases characterized by a set of $\delta+1$ topological invariants, ranging from the $D$-dimensional topological invariant to the $(D-\delta)$-dimensional topological invariant. The system exhibits topological response signatures and bulk-boundary correspondences governed by combinations of these topological invariants taking non-trivial values, with lower-dimensional topological invariants characterizing fragmentation of the underlying topological phase of the system thermodynamically large in all $D$-dimensions. We demonstrate this physics for the paradigmatic Chern insulator phase, but show its requirements for realization are satisfied by a much broader set of topological systems.
\end{abstract}

\maketitle

Consequences of topology in condensed matter physics frequently stem from the non-trivial topological invariant of a material bulk associated with incompressible states, corresponding to topological response signatures of the bulk and topologically-protected boundary states~\cite{Laughlin_1981, Halperin_1982, Hasan2011}. These signatures yield the unpaired Majorana zero-modes required for topological quantum computation~\cite{Kitaev_2001, Read_2000, Ladd_2010, Raussendorf_2007, Freedman2003,nayak_2008, Kitaev_2003, Alicea_2016} and topological boundary metals useful for spintronics devices~\cite{Fert1988, Gruenberg1988, Murakami_2003, Sinova_2004, Koenig_2008,  Lesne_2016, Kondou_2016, Bernevig_2006, Koenig_2008}, for instance, reflecting their significance. While the relationship between bulk topological responses and topological boundary states has typically been studied in systems large enough that finite-size effects are neglected while still yielding good agreement with experiment~\cite{Ryu_2010, Chiu2016, Schnyder2008}, significant improvements in fabrication techniques now permit experimental study of lower-dimensional systems in which finite-size effects are significant. Past work on such systems reports findings of lower-dimensional topological phases~\cite{Potter2010,Asmar2018,He2018,Otrokov2019,Chowdhury2019,Li2019,Lei2020,Liu2021} or higher-dimensional phases~\cite{Chowdhury2019,Gong2019,Irfan2020,Mir2020,Thalmeier2020,Xu2020,Zhao2021,Asmar2022} rather than the transition from higher-dimensional to lower-dimensional topological phases itself, motivating greater scrutiny of this process.

We address the need for greater understanding of this transition between higher-dimensional and lower-dimensional topological phases by showing that finite-size effects can yield \textit{additional, previously-unidentified topological phases}. In the case of the Chern insulator, which is two-dimensional (2D) in the bulk, we consider opening boundary conditions in one direction, and thin the system in this direction to a quasi-one-dimensional geometry, which we denote as quasi-(2-1)D as the underlying bulk is 2D. While dimensional reduction also considers opening boundary conditions in one direction and thinning the system in one direction, the system size is large in the direction in which boundary conditions are opened, and small in the periodic direction~\cite{Qi2008_fieldtheory}. Dimensional reduction also holds even when the system is thinned to be strictly one-dimensional. We note that this scenario considered for dimensional reduction, of system size large in the direction of open boundary conditions, is also the regime considered by the ten-fold way classification scheme~\cite{Ryu_2010}. We instead consider thinning the system in the direction in which boundary conditions are opened, to finite thicknesses small relative to the penetration depth of topological boundary modes, and keep system size thermodynamically large in the periodic direction.

We also report on effects here, which only occur for systems finite in width in the direction of open boundary conditions. In this case, the quasi-(2-1)D Chern insulator can still exhibit charge pumping in response to changing magnetic flux through a plaquet of the lattice due to non-trivial Chern number. However, if we open boundary conditions in the second direction, the quasi-(2-1)D Chern insulator also exhibits a second bulk-boundary correspondence, with quasi-zero-dimensional---or quasi-(2-2)D---topologically-protected, gapless boundary modes localized at the ends of the quasi-(2-1)D system, in correspondence with a lower-dimensional topological invariant characterizing topology of the quasi-(2-1)D bulk. These quasi-zero-dimensional states are topologically robust and cannot be understood purely through interference of chiral modes. Furthermore, the number of topologically-robust quasi-zero-dimensional modes at each end of the ribbon can be integer-valued in direct correspondence with the Chern number, in the presence of only a $\sigma_z \mathcal{I}$ symmetry.  This symmetry may be interpreted as spatial inversion symmetry depending upon the choice of physical degrees of freedom, but this integer classification indicates that $\mathbb{Z}_2$ topological invariants of 1D topological insulators protected by spatial inversion symmetry are therefore unsuitable~\cite{Hughes2011}. Instead, $\mathbb{Z}$ classification indicates previously-unidentified topological phases characterized by both a $\mathbb{Z}$ Chern number and a related $\mathbb{Z}$ lower-dimensional topological invariant. As topological phases are defined by their topological invariant(s)~\cite{Schnyder2008,Ryu_2010, niu1985}, the characterization of topological states of finite-size systems in terms of a set of topological invariants of different dimensionalities---and in terms of more topological invariants than required for $D$-dimensional systems which are thermodynamically large in all $D$ dimensions---is fundamentally different from previous work.

We consider finite-size topological phases here for systems with a $\sigma_z \mathcal{I}$ symmetry, where $\mathcal{I}$ is lattice inversion and $\sigma_z$ acts on the orbital space and squares to the identity. Our results are therefore broadly applicable to topological phases irrespective of other symmetries present and bulk dimensionality, in particular given the Chern insulator phase is used to construct many other topological phases of matter~\cite{Verresen2017, Bernevig_2006, Hasan2011, Moore_2007, Lindner_2011, Shen_2018, Kane_2014, Ozawa_2019, Wu_2017, Onose_2010}. We expect finite-size topological phases to be prominent in lower-dimensional materials with non-trivial topology such as stacked van der Waals materials and nanowires.

\textit{Finite-size topology of the Chern insulator}---We first show that finite-size topology occurs in the 2D Chern insulator phase. The phase is foundational both in understanding the quantum anomalous Hall effect~\cite{liu_2016} and as a building block used to construct many other topological phases of matter, most notably the quantum spin Hall insulator~\cite{kane_2005, Bernevig_2006}, the three-dimensional topological insulator~\cite{Moore_2007}, and the Weyl semimetal~\cite{wan_2011, burkov_2011}. Various analogues of the Chern insulator that are not electronic~\cite{mcclarty_2021, Ma_2019} and/or out of equilibrium~\cite{harper_2020, Gong_2018} are also expected to display finite-size topology.

The considered Chern insulator Hamiltonian
\begin{align}\label{HC1}
    &H_{\textrm{CI}} = \sum_{\sigma,n,m} \sigma (M+\kappa_{n,m}) c_{\sigma,n,m}^{\dag}c_{\sigma,n,m}\\
    &-t\sum_{\sigma,n,m}  (\sigma c_{\sigma,n,m}^{\dag}c_{\sigma,n+1,m}+(\sigma-\epsilon)c_{\sigma,n,m}^{\dag}c_{\sigma,n,m+1}+h.c.)\nonumber\\
    &-\Delta\sum_{\sigma,n,m} (\sigma c_{\sigma,n,m}^{\dag}c_{-\sigma,n+1,m} +ic_{\sigma,n,m}^{\dag}c_{-\sigma,n,m+1}+h.c.),\nonumber
\end{align}
is a simplified version of the Qi-Wu-Zhang model~\cite{qi2006_QWZmodel} on an $L_x\times L_y$ square lattice, where $c^\dag_{\sigma,n,m}$ creates a particle in the orbital $\sigma\in\{-1,1\}$ at site position $(n,m)$, $M$ is the Zeeman field strength, $t$ is the strength of real hopping in both $\hat{x}$- and $\hat{y}$-direction without changing the orbital, and $\Delta$ is the strength of the hopping term changing the orbital. For $\epsilon=0$ this Hamiltonian has a particle-hole symmetry described by the operator $\sigma_x K$, and for $\kappa_{n,m}=0$ it has $\sigma_z \mathcal{I}$ symmetry. Here, $\sigma_x$ and $\sigma_z$ are Pauli matrices acting on the orbital space, $K$ is complex conjugation, and $\mathcal{I}$ is inversion of the 2D lattice. Unless specified otherwise, we take $t=1$, $\Delta=0.22$, $\epsilon=\kappa_{n,m}=0$, and $L_y=600$ in the numerical computations below. For these parameters, the Hamiltonian realizes a Chern insulator phase with Chern number $\mathcal{C}=+1$ for $-4<M<0$ and a distinct Chern insulator phase with Chern number $\mathcal{C}=-1$ for $0<M<4$. In the following, we use (pe,pe), (op,pe), and (op,op) to refer to different combinations of open (op) and periodic (pe) boundary conditions in the $(\hat{x},\hat{y})$-directions.

\begin{figure}
    \includegraphics[width=\linewidth]{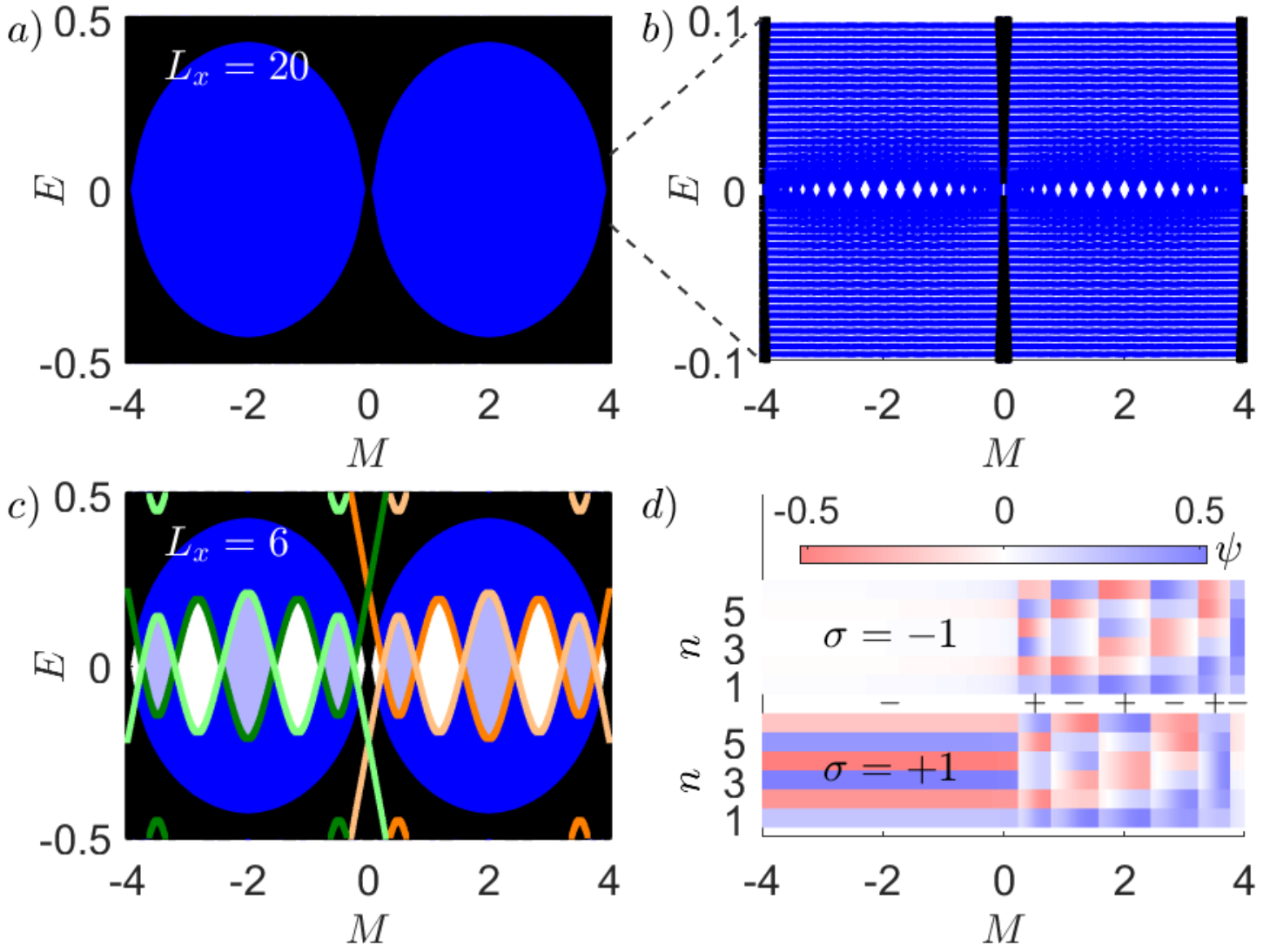}
    \caption{(a) The single-particle energy spectrum $E$ of the considered Chern insulator model on a $20\times600$ square lattice with (pe,pe) boundary conditions (black) is gapped except at a few values of the Zeeman field strength $M$. The spectrum for (op,pe) boundary conditions (blue) has states inside this gap. (b) When zooming (a) to a smaller energy range, small gaps in the (op,pe) spectrum become discernible. (c) When $L_x$ is decreased, the small gaps become larger. Eigenstates in the (op,pe) spectrum with momentum $k_y=0$ ($k_y=\pi$) are highlighted by their $\sigma_z \mathcal{I}$ eigenvalue (light orange (green) for $+1$ eigenvalue and dark orange (green) for $-1$ eigenvalue). Intervals in $M$ over which the highest-energy occupied state at half filling within the $k_y=0$ and $k_y=\pi$ sectors have opposite $\sigma_z \mathcal{I}$ eigenvalue are highlighted with purple background color. (d) Components of the highest-energy occupied state within the $k_y=0$ sector versus $M$ for $L_x=6$ and (op,pe) boundary conditions. The wavefunction is real, and the components are labeled by orbital $\sigma$ and unit cell position $n$ in the $\hat{x}$-direction. The $\sigma_z \mathcal{I}$ eigenvalues are indicated with plus and minus signs.}
    \label{fig:spectrumbubbles}
\end{figure}

\begin{figure}
    \includegraphics[width=\linewidth]{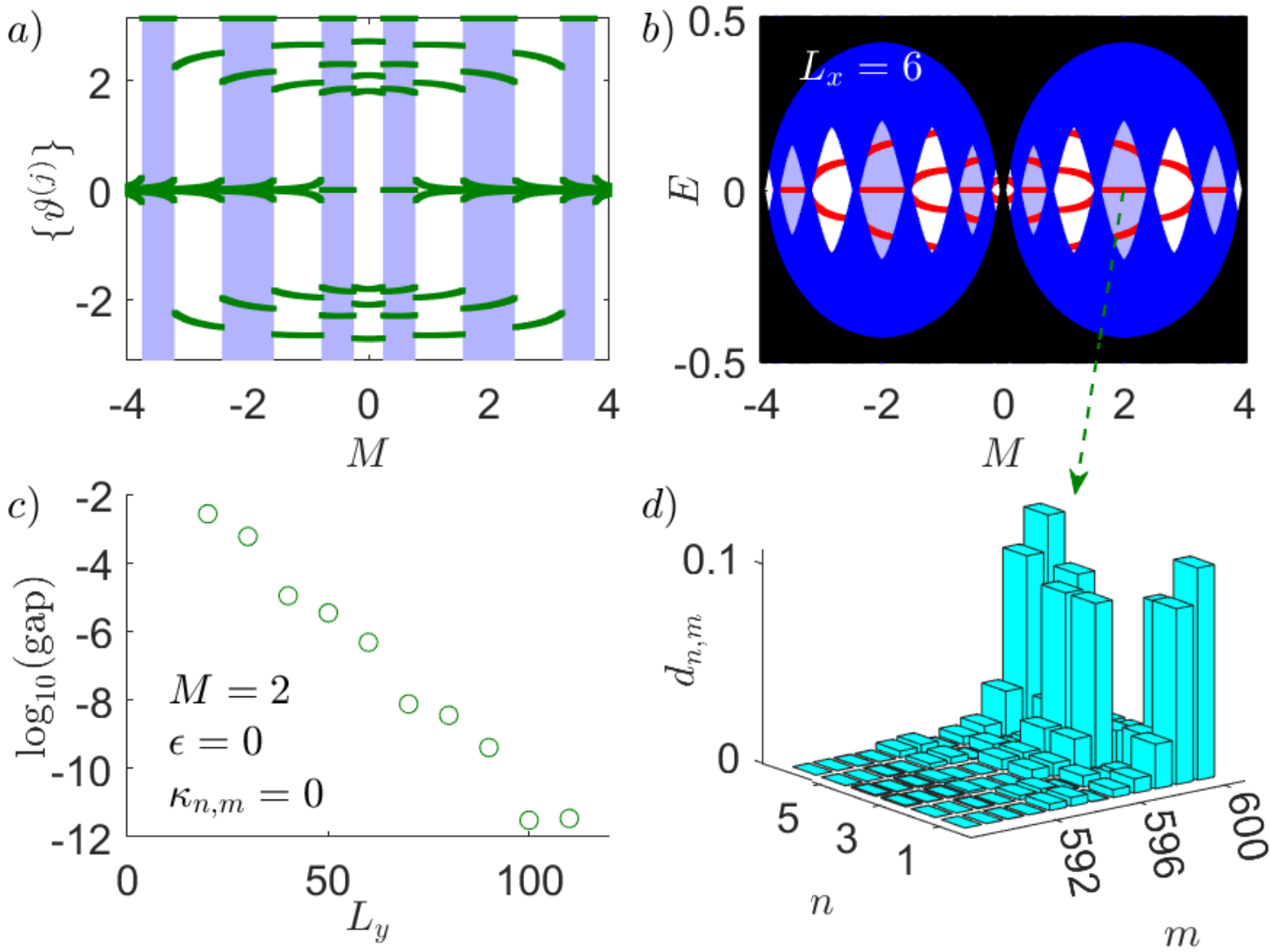}
    \caption{1D topology in the quasi-(2-1)D system. (a) Wannier spectrum versus Zeeman field strength $M$ for (op,pe) boundary conditions and $L_x = 6$. Intervals in $M$, for which one of the Wannier charge centers is $\pm\pi$, are highlighted in purple. (b) Results of Fig.~\ref{fig:spectrumbubbles}(c) for (pe,pe) and (op,pe) boundary conditions superimposed over the corresponding (op,op) spectrum (red). (c) The energy gap between the nearly degenerate in-gap states at zero energy in the (op,op) spectrum decreases exponentially with $L_y$, with variations due to Friedel oscillations. (d) Density distribution $d_{n,m}=\sum_\sigma\langle c^\dag_{\sigma,n,m} c_{\sigma,n,m}\rangle$ of one of the two states at zero energy in the quasi-(2-1)D bulk gap for (op,op) boundary conditions and $M=2$. We show only the outermost 12 rows of sites for the slab as more than $99.8\%$ of the total density is located here. The other state is localized at the other end.}
    \label{fig:q1Dbulkboundary}
\end{figure}

\begin{figure}
    \includegraphics[width=\linewidth]{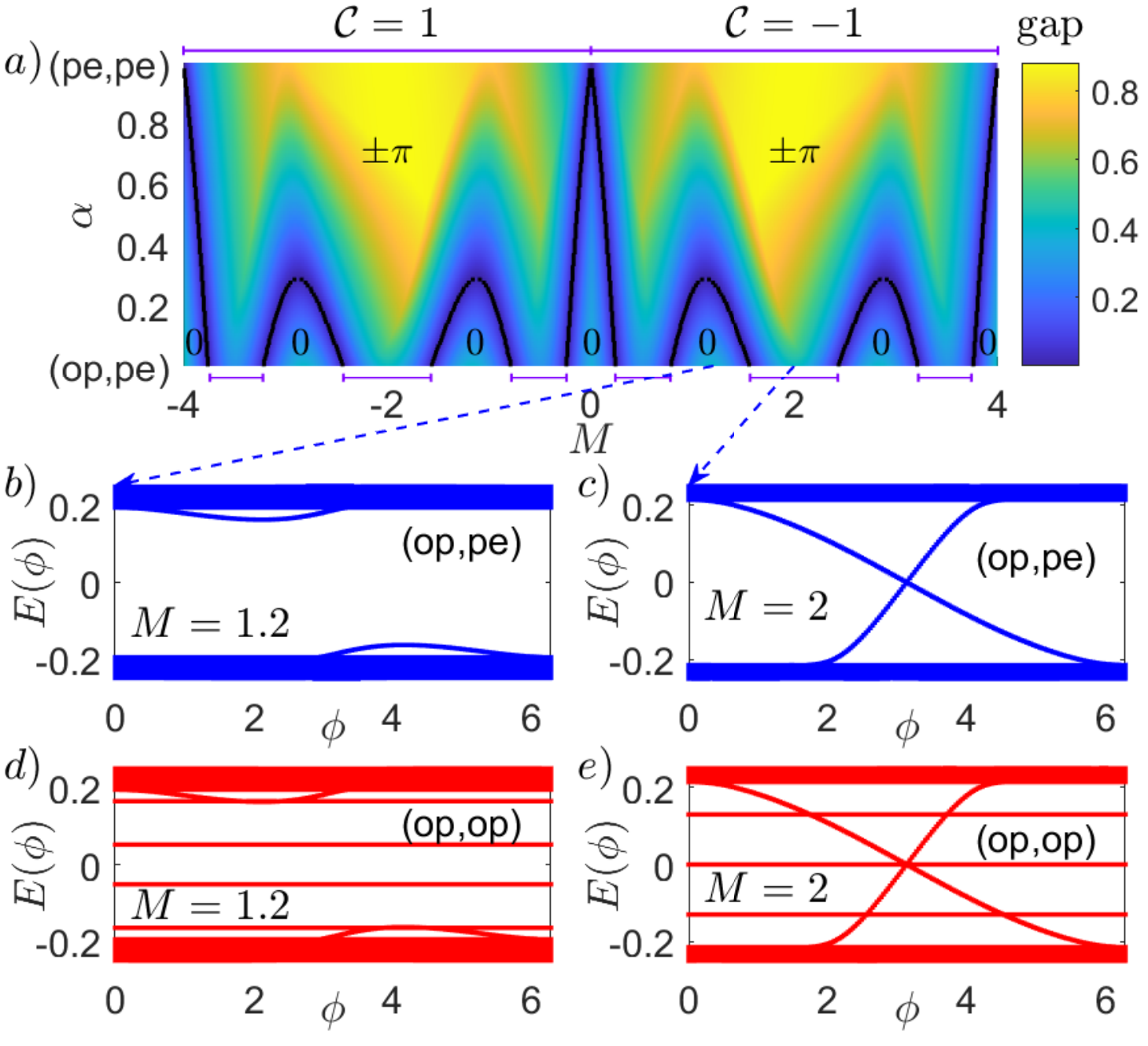}
    \caption{2D topological response in the quasi-(2-1)D system. (a) We continuously open the boundary conditions in the $\hat{x}$ direction by scaling the hopping strengths of all hops across the boundary by $\alpha$. In this process, gap closings happen for some $M$ but not for others. The regions where gap closings do not occur remain topological. The black lines that fall on top of the gap closings are the boundaries between topological regions in which the Wannier charge center of the highest-energy occupied state is $\pm\pi$ and trivial regions in which it is $0$. The purple line segments above and below the plot show the topological regions for (pe,pe) and (op,pe) boundary conditions, respectively. (b,c) When inserting a flux $\phi$ through the central plaquette of the $6\times600$ lattice, the single-particle spectrum of the Chern insulator model shows (b) no pumping inside trivial bubbles and (c) Thouless pumping inside topological bubbles across the gap in the (op,pe) spectrum. (d,e) The in-gap states for (op,op) boundary conditions are not affected by the flux insertion, as they are localized at the ends of the slab.}
    \label{fig:flux}
\end{figure}

\begin{figure}
    \includegraphics[width=\linewidth]{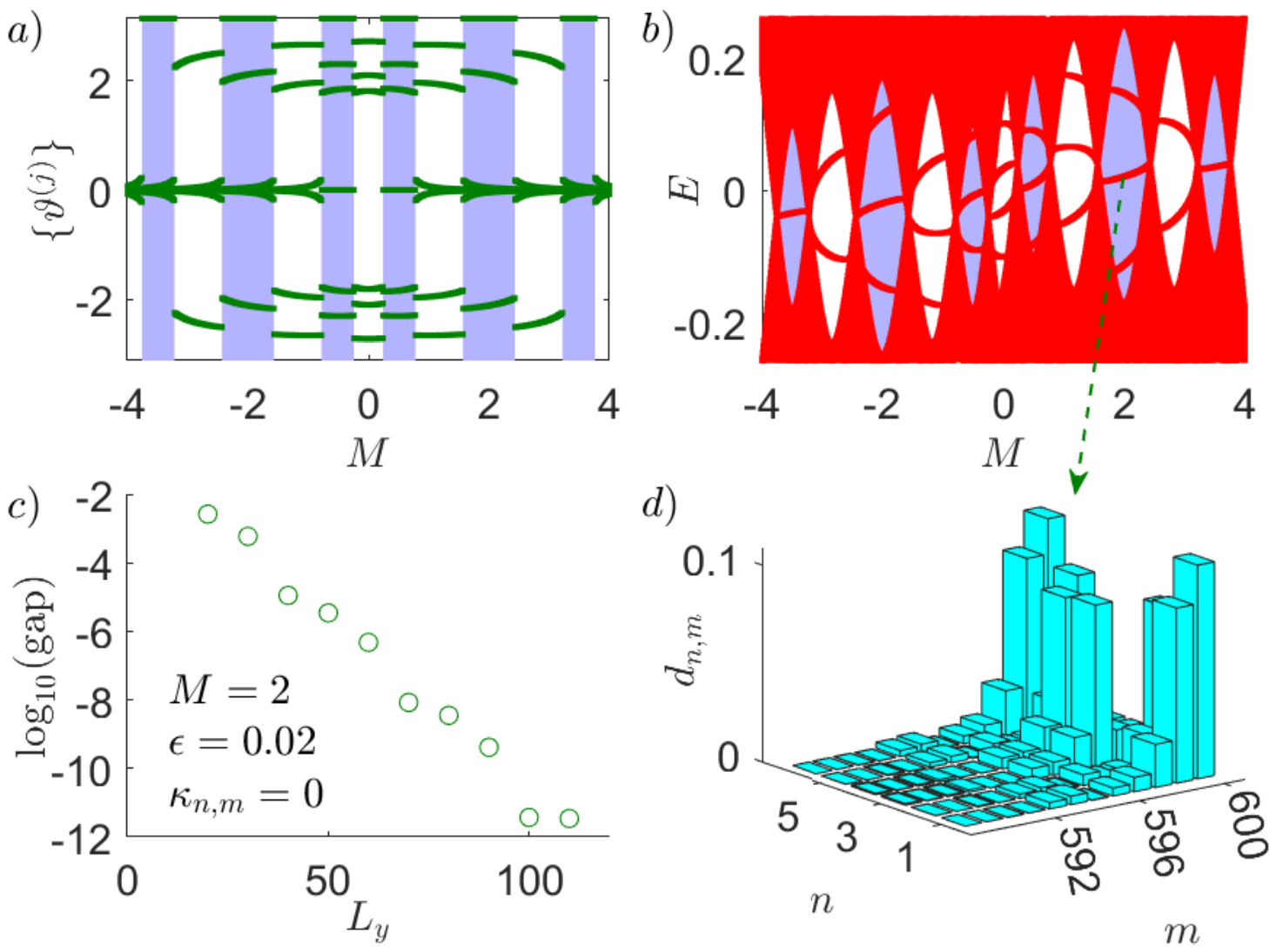}
    \caption{(a) Wannier spectrum versus Zeeman field strength $M$ when particle-hole symmetry is broken by taking $\epsilon=0.02$. $M$ intervals for which at least one Wannier charge center is $\pm\pi$ are highlighted in purple. (b) Single-particle spectrum for (op,op) boundary conditions (red) of the Chern insulator slab with $L_x=6$, $L_y=600$, $\epsilon=0.02$, and $\kappa=0.02$ averaged over $200$ disorder realizations (see Supplemental Material, Sec.\ III, for details of disorder treatment). The in-gap states are two-fold degenerate, and the purple background shows the topological regions computed for $\kappa=0$. (c) The energy gap between the nearly-degenerate, quasi-(2-2)D boundary modes (states $L_xL_y$ and $L_xL_y+1$) for $M=2$, $\epsilon=0.02$, and $\kappa=0$ decreases exponentially with $L_y$. (d) The quasi-(2-2)D nature of the boundary modes inside the bubble gaps is manifested in the density distributions $d_{n,m}=\sum_\sigma\langle c^\dag_{\sigma,n,m} c_{\sigma,n,m}\rangle$ of the states, here plotted for $M=2$, $\epsilon=0.02$, $\kappa=0.02$, and one disorder realization. We show only the outermost
12 rows of sites for the slab as more than $99.8\%$ of the total density is located here. The other state is localized at the other end. We obtain similar results for $L_x=5$ (see Supplemental Material, Secs.\ II and III).}
    \label{fig:density}
\end{figure}

For a gapped topological phase characterized in a 2D bulk by a non-trivial Chern number $\mathcal{C}$, we generally expect $\mathcal{C}$ chiral modes localized on each edge, if the boundary conditions are opened in one direction $\hat{x}$ and the system size in the $\hat{x}$ direction, $L_x$, is sufficiently large. The edge modes give rise to the blue states inside the 2D bulk energy gap seen in Fig.\ \ref{fig:spectrumbubbles}(a). As $L_x$ is decreased, however, the system becomes quasi-(2-1)D. The states appearing within the 2D bulk energy gap increasingly permeate into the quasi-(2-1)D bulk and hybridize, such that we can only refer to them as states within the 2D bulk gap resulting from bulk-boundary correspondence, rather than edge states. The hybridization among these states can produce gaps in the quasi-(2-1)D spectrum. In fact, small gaps can already be seen for $L_x=20$ (Fig.\ \ref{fig:spectrumbubbles}(b)), but the effect is much larger for smaller $L_x$ (Fig.\ \ref{fig:spectrumbubbles}(c)). When viewed as a function of a parameter of the model, the gaps can form bubbles separated by transition points as seen in Fig.\ \ref{fig:spectrumbubbles}(c). A general mechanism to obtain such bubbles is when the highest-energy occupied state and lowest-energy unoccupied state of the quasi-(2-1)D bulk belong to different symmetry sectors and oscillate out of phase with sufficiently large amplitude as a function of the considered parameter. We will describe below how this happens for the Chern insulator. For the quasi-(2-1)D Chern insulator, we find that there are $2L_x$ transition points and hence $2L_x-1$ bubbles.

We first consider the case $\kappa_{n,m}=\epsilon=0$. If $|\psi\rangle$ is an eigenstate of $H_{\textrm{CI}}$ with energy $E$ and an eigenstate of $\sigma_z\mathcal{I}$ with eigenvalue $s$, then $\sigma_x K|\psi\rangle$ is an eigenstate of $H_{\textrm{CI}}$ with energy $-E$ and an eigenstate of $\sigma_z\mathcal{I}$ with eigenvalue $-s$. Non-degenerate pairs of states with energies $\pm E$ hence have opposite $\sigma_z\mathcal{I}$ eigenvalues. Note also that $H_{\textrm{CI}}$ and $\sigma_z\mathcal{I}$ can be simultaneously diagonalized within the sector with momentum $k_y=0$ in the $\hat{y}$-direction. The $\sigma=+1$ component of the resulting eigenstates is symmetric (antisymmetric) under the mirror operation that takes site $n$ into $L_x+1-n$ if the $\sigma_z\mathcal{I}$ eigenvalue is $+1$ ($-1$) while the $\sigma=-1$ component is antisymmetric (symmetric). Similar considerations apply for $k_y=\pi$.

We observe numerically (Fig.\ \ref{fig:spectrumbubbles}(c)) that the lowest-energy unoccupied state and the highest-energy occupied state forming the $L_x-1$ bubbles at $M>0$ have momentum $k_y=0$. Plotting the highest-energy occupied state at half filling within the $k_y=0$ sector (Fig.\ \ref{fig:spectrumbubbles}(d)), we observe that the wavelength of the wavefunction in the $\hat{x}$ direction decreases as $M$ goes from $4$ to $0$. Let us consider the $\sigma=-1$ components of the state. The global phase of the wavefunction is chosen such that the component at $n=1$ is positive. For symmetric (antisymmetric) states, the component at $n=L_x$ is hence positive (negative). A smoother change in the wavelength as a function of $M$ can hence be obtained by alternating between symmetric and antisymmetric states, and this alternation produces the bubbles. The alternating sign of the $\sigma_z\mathcal{I}$ eigenvalue for the highest-energy occupied state also means that the product of $\sigma_z \mathcal{I}$ eigenvalues over high-symmetry points in the slab Brillouin zone undergoes a relative change in sign when tuning $M$ through a transition point between two bubbles. The transition points therefore generically correspond to topological phase transitions.

We may further characterize the topology of the bubbles as topologically non-trivial or trivial by considering half filling and computing the eigenvalue spectrum of the (discretized) Wilson loop operator \cite{WLpaper}
\begin{equation}\label{Wilsoneqn}
W_{jk}=\lim_{S\to\infty}\langle \psi_{0}^{(j)}| P_{S-1}P_{S-2}\cdots P_{2} P_{1}|\psi_{0}^{(k)}\rangle,
\end{equation}
where $|\psi_{0}^{(j)}\rangle$ is the $j$th occupied energy eigenstate within the $k_y=0$ sector and $P_{l}$ is the projector onto the occupied subspace within the $k_{y}=2\pi l/S$ momentum sector. The eigenvalues of $W_{jk}$ are phase factors $\{ e^{i\vartheta^{(j)}} \}$, and the model is topological if at least one of the Wannier charge centers $ \{\vartheta^{(j)} \}$ is $\pm\pi$. Such characterization of the quasi-(2-1)D bulk topology is depicted in Fig.\ \ref{fig:q1Dbulkboundary}(a).

We observe numerically that the highest-energy occupied state as a function of $k_y$ is separated from the remaining occupied states (see Supplemental Material, Sec.\ I), and hence we can also compute the Wannier charge center for this state alone. For systems with inversion symmetry, one can determine the number of Wannier charge centers that are $\pm\pi$ from the inversion eigenvalues at the momenta $k_y=0$ and $k_y=\pi$ \cite{WLpaper}. This explains the agreement between the purple regions in Figs.\ \ref{fig:q1Dbulkboundary}(a) and \ref{fig:spectrumbubbles}(c).

There is \textit{an additional bulk-boundary correspondence} of the quasi-(2-1)D Chern insulator when the Wilson loop spectrum possesses topologically non-trivial eigenvalues: when $M$ lies inside an interval corresponding to a topological bubble for open boundary conditions in the $\hat{x}$-direction and finite $L_x$, additionally opening boundary conditions in the $\hat{y}$-direction yields a pair of topological, quasi-(2-2)D gapless boundary modes at zero energy inside the bubble (Fig.~\ref{fig:q1Dbulkboundary}(b)). The energy gap between these two states displays exponential decay to zero with increasing $L_y$ (Fig.~\ref{fig:q1Dbulkboundary}(c)), and the probability density of one of these states for (op,op) boundary conditions is shown in Fig.~\ref{fig:q1Dbulkboundary}(d). The probability density is strongly-localized at one end of the system in the $\hat{y}$-direction, such that it is quasi-0D. As the system with (pe,pe) boundary conditions is two-dimensional, we more precisely identify these states as quasi-(2-2)D. The finite-size topology in the Chern insulator has $\mathbb{Z}$ topological classification, as discussed in the Supplemental Material, Sec.~IV.

While there are additional states at non-zero energy in the bubbles for the parameter sets shown and they are localized and robust against disorder while in the bubble gap, these states are consumed by the quasi-(2-1)D bulk through smooth deformation of the system that reduces the maximum height of the bubbles in energy. The quasi-(2-2)D states at nearly zero energy, in contrast, occur in correspondence with the non-trivial Wilson loop spectrum of these bubbles, and are only removed by closing the quasi-(2-1)D bulk gap.

In addition to the bulk-boundary correspondence between a topological invariant of the quasi-(2-1)D bulk and quasi-(2-2)D topologically-protected boundary states, however, the finite-size topology Chern insulator is adiabatically connected to a 2D system with nonzero Chern number (Fig.\ \ref{fig:flux}(a)), and response signatures of this topology persist in the quasi-(2-1)D system, clearly distinguishing a quasi-(2-1)D finite-size topological phase from a 1D topological phase. This is demonstrated by computing the evolution of the spectrum for the quasi-1D Chern insulator with open boundary conditions in each direction, as a function of magnetic flux $\phi$ through the center plaquette of the lattice (Fig.~\ref{fig:flux}(b-e)). Such $\phi$-dependence of the energy spectrum results from the dependence of charge density on applied magnetic field strength determined by the Chern number~\cite{KTChen2011}. We find such charge pumping in the spectrum versus $\phi$, but \textit{only for topologically non-trivial bubbles}. Notably, the quasi-(2-2)D boundary states remain at fixed energy while $\phi$ is varied, reflecting their dependence on the topologically non-trivial polarization invariant of the quasi-(2-1)D bulk rather than on the full Chern number.

We also consider the Chern insulator with a finite-width wire geometry in the case of the particle-hole symmetry-breaking term $\epsilon$ being nonzero. Also in this case, we observe topological and trivial bubbles in the spectrum as a function of $M$ characterized by quantized Wilson loop eigenvalues (Fig.\ \ref{fig:density}(a)) and pairs of degenerate quasi-(2-2)D in-gap states localized at the ends of the wire (Fig.\ \ref{fig:density}(c)). We also observe flux pumping similar to the results in Fig.\ \ref{fig:flux}. If we add disorder of strength $\kappa=0.02$ by randomly choosing $\kappa_{n,m}\in[-\kappa,\kappa]$, both the momentum and the $\sigma_z \mathcal{I}$ symmetry are broken. The (op,op) spectrum still shows bubbles with in-gap states as a function of $M$. The in-gap states are localized at the ends of the wire (Fig.\ \ref{fig:density}(d)), but there is now an energy gap between pairs of in-gap states with densities at opposite ends. If the spectrum is averaged over several disorder realizations, however, this gap averages to zero (Fig.\ \ref{fig:density}(b) and Supplemental Material Sec.\ III).

\textit{Conclusion}---We show $D$-dimensional topological phases exhibit additional non-trivial \textit{finite-size} topology:  topologically-protected $(D-1)$-dimensional boundary modes resulting from  non-trivial topology of a $D$-dimensional bulk can hybridize in finite-size systems to induce additional topological phase transitions in the finite $D$-dimensional system with open boundary conditions \textit{even when the $D$-dimensional bulk gap remains open}. This additional non-trivial topology in the finite-size $D$-dimensional system can yield an additional bulk-boundary correspondence to realize additional topologically-protected quasi-($D$-2)-dimensional boundary states, while the system still exhibits response theory of a $D$-dimensional topological invariant. We show such finite-size topology occurs in Chern insulators, realizing quasi-(2-2)D topological modes in a quasi-(2-1)D system and charge pumping in response to an applied magnetic field. As Chern insulators are the basis for many other models and topological phases~\cite{Verresen2017, Bernevig_2006, Hasan2011, Moore_2007, Lindner_2011, Shen_2018, Kane_2014, Ozawa_2019, Wu_2017, Onose_2010}, such finite-size topology is a generic property that necessitates reexamination of known topological phases, to be explored in future work.

\begin{acknowledgments}

\textit{Acknowledgements}---We thank Joel E. Moore for helpful discussions. This work has been supported by Danmarks Frie Forskningsfond under Grant No.\ 8049-00074B and Carlsbergfondet under Grant No.\ CF20-0658. This research was also supported in part by the National Science Foundation under Grant No.\ NSF PHY-1748958.

\textit{Author Contributions}---A.~M.~C.\ developed the concept. A.~M.~C.\ and A.~E.~B.~N.\ performed calculations, interpreted results, and wrote the manuscript.

\end{acknowledgments}

\appendix

\section{Supplemental Material I. Dispersion relations for the Chern insulator}

\begin{figure*}[t]
\includegraphics[trim=96 240 98 255,clip, width=0.33\linewidth]{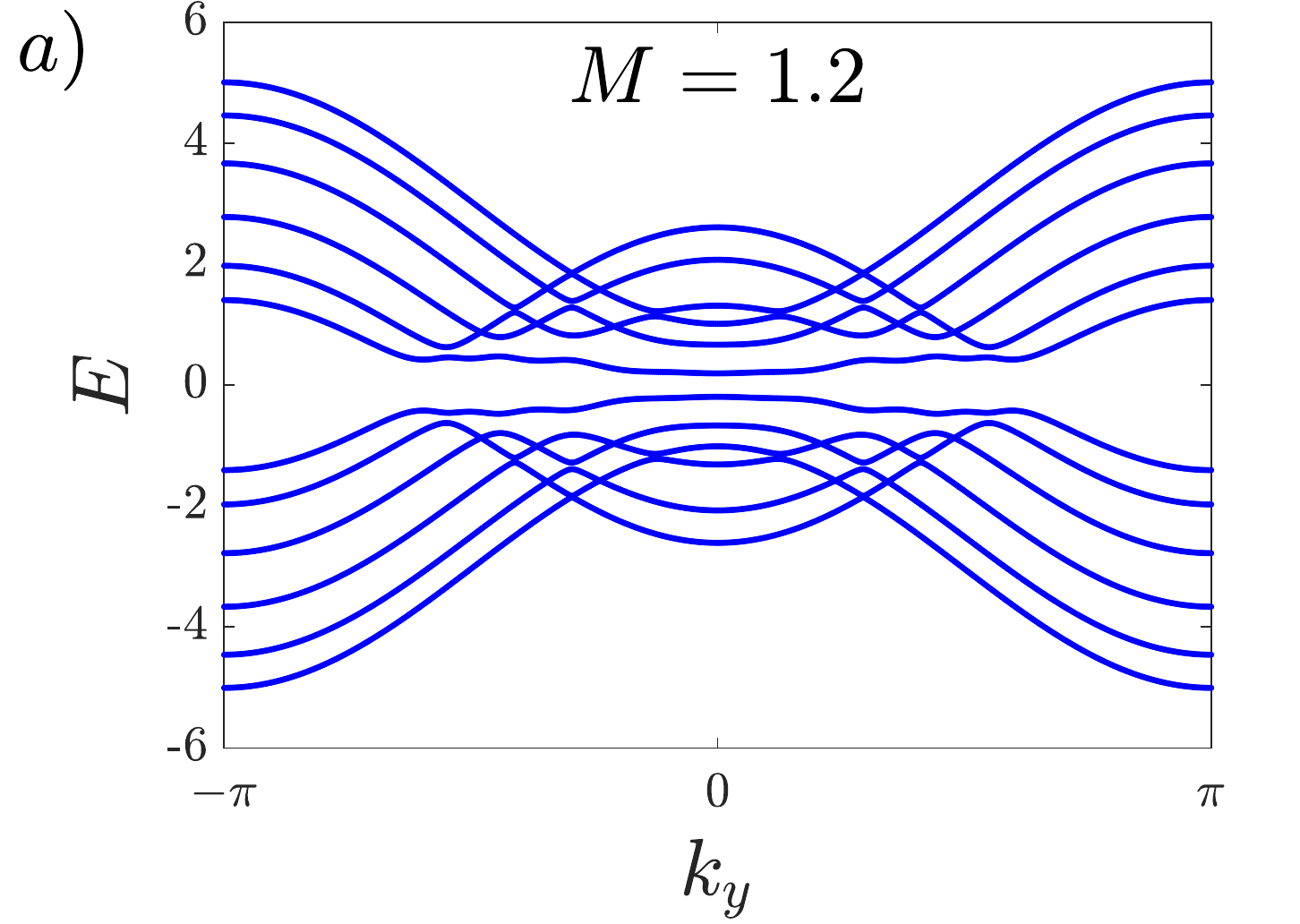}\hfill
\includegraphics[trim=96 240 98 255,clip, width=0.33\linewidth]{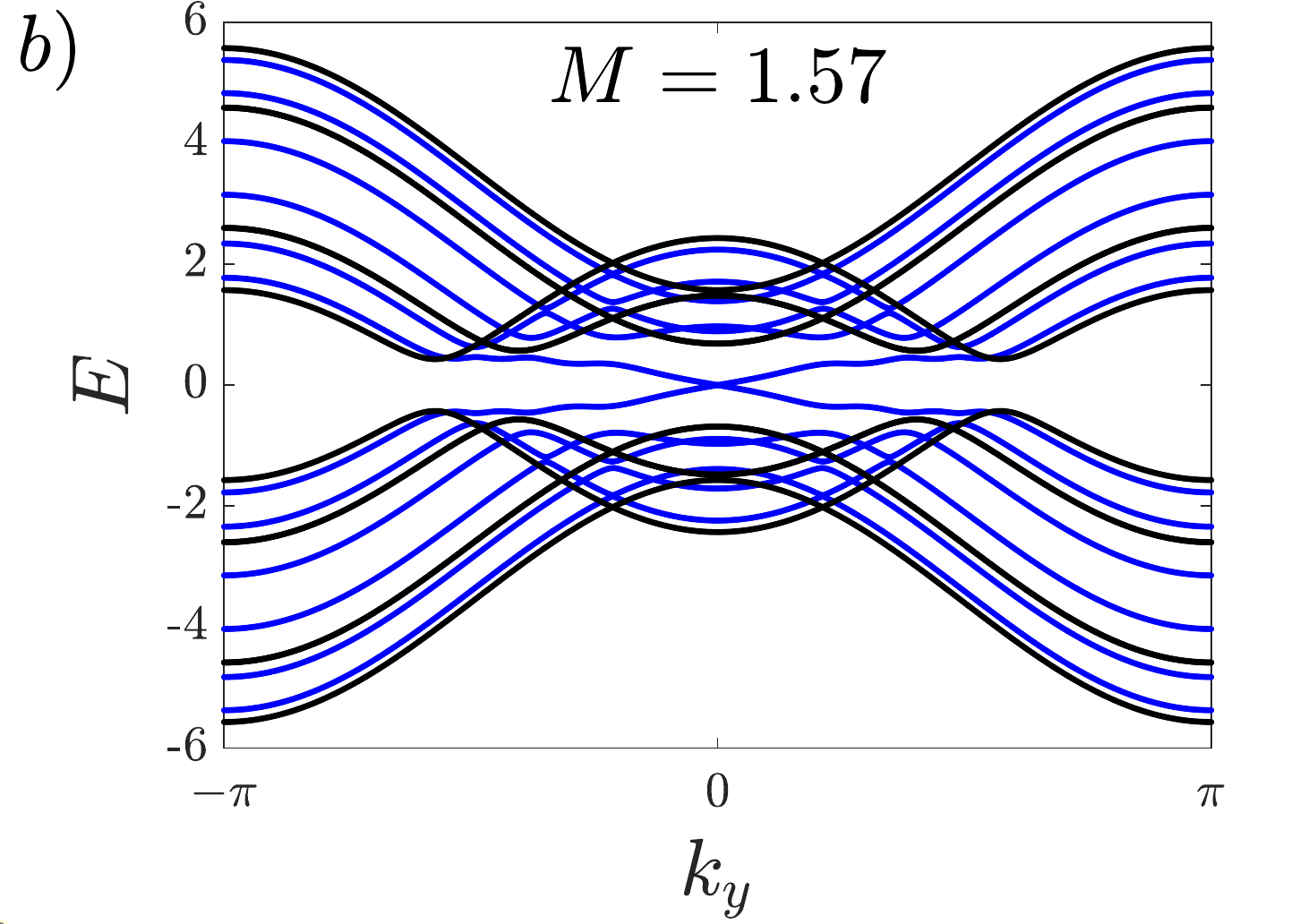}\hfill
\includegraphics[trim=96 240 98 255,clip, width=0.33\linewidth]{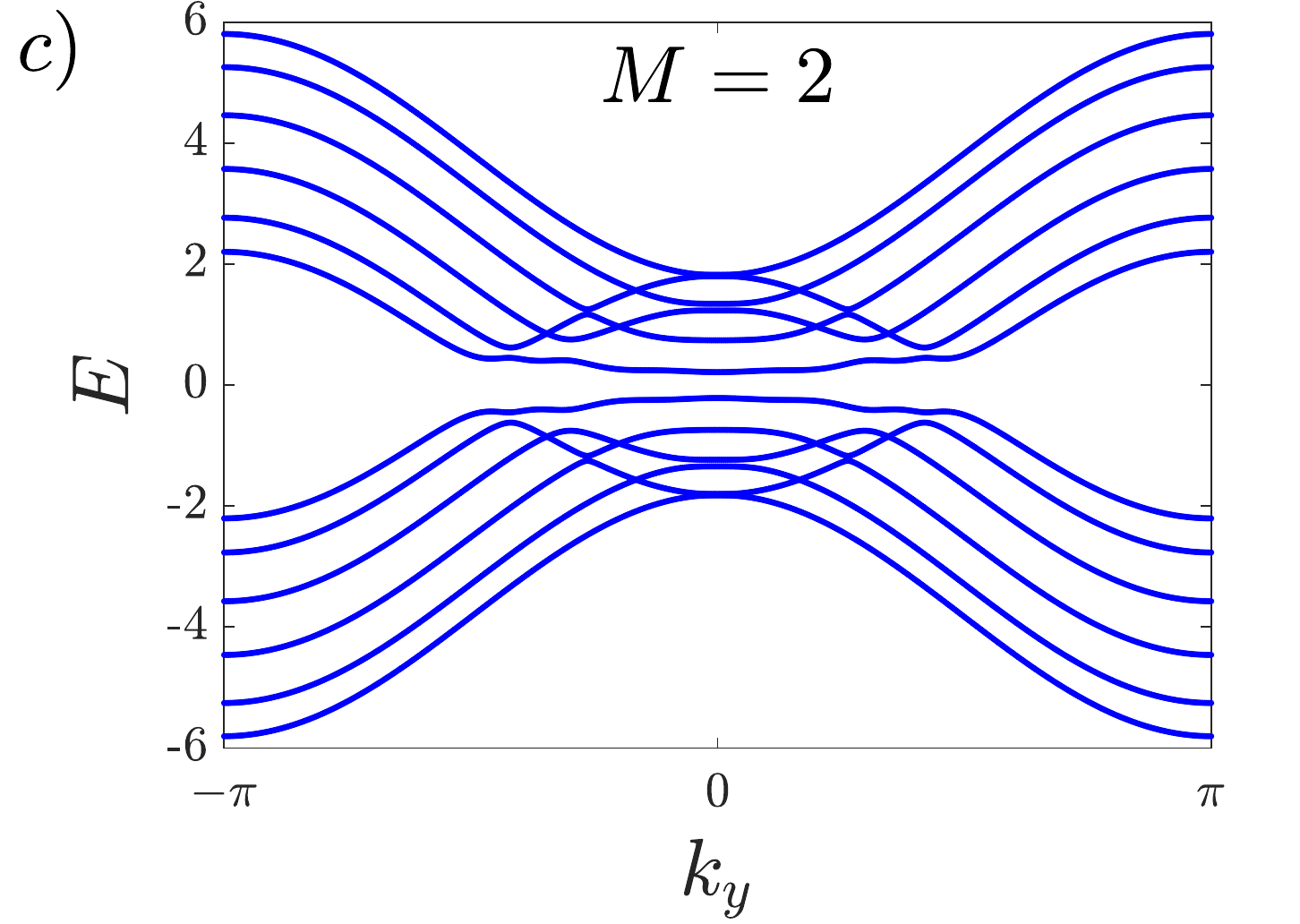}
\caption{Dispersion relations for the Chern insulator model in Eq.~(1) in the main text with $L_x=6$, $t=1$, $\Delta=0.22$, $\epsilon=0$, $\kappa_{n,m}=0$, and (op,pe) boundary conditions (blue) for different values of $M$. The plot for $M=1.57$ also shows the single-particle spectrum for (pe,pe) boundary conditions (black).
For $M=1.2$ the system is inside a trivial bubble, and for $M=2$ the system is inside a toplogical bubble. The plot at $M=1.57$ shows the gap closing in the (op,pe) spectrum that happens at the phase transition between the trivial and the topological bubble. This gap closing happens while the gap in the (pe,pe) spectrum remains open. Note also that the highest state below zero energy is separated from all the other states below zero energy. We can therefore compute the Wannier charge center for this state alone. The result is $0$ for $M=1.2$ and $\pi$ for $M=2$. This shows that the topological information is carried by this state.}
\label{fig:many1}
\end{figure*}

In Fig.~\ref{fig:many1}, we show dispersion relations for the Chern insulator model with $L_x=6$. For open boundary conditions in the $x$ direction, the chiral modes of the $M>0$ Chern insulator cross at the center of the slab Brillouin zone, as shown in Fig.~\ref{fig:many1}(b). Chiral boundary modes of the Chern insulator phase for $M<0$ instead cross at the edge of the slab Brillouin zone. The figure also shows that the highest energy state below zero energy is separated from the other states below zero energy, which allows us to compute the Wannier charge center for this state alone.

\section{Supplemental Material II. Finite-size topology in the Chern insulator with $L_x=5$}

\begin{figure*}
\includegraphics[trim=96 240 98 255,clip, width=0.33\linewidth]{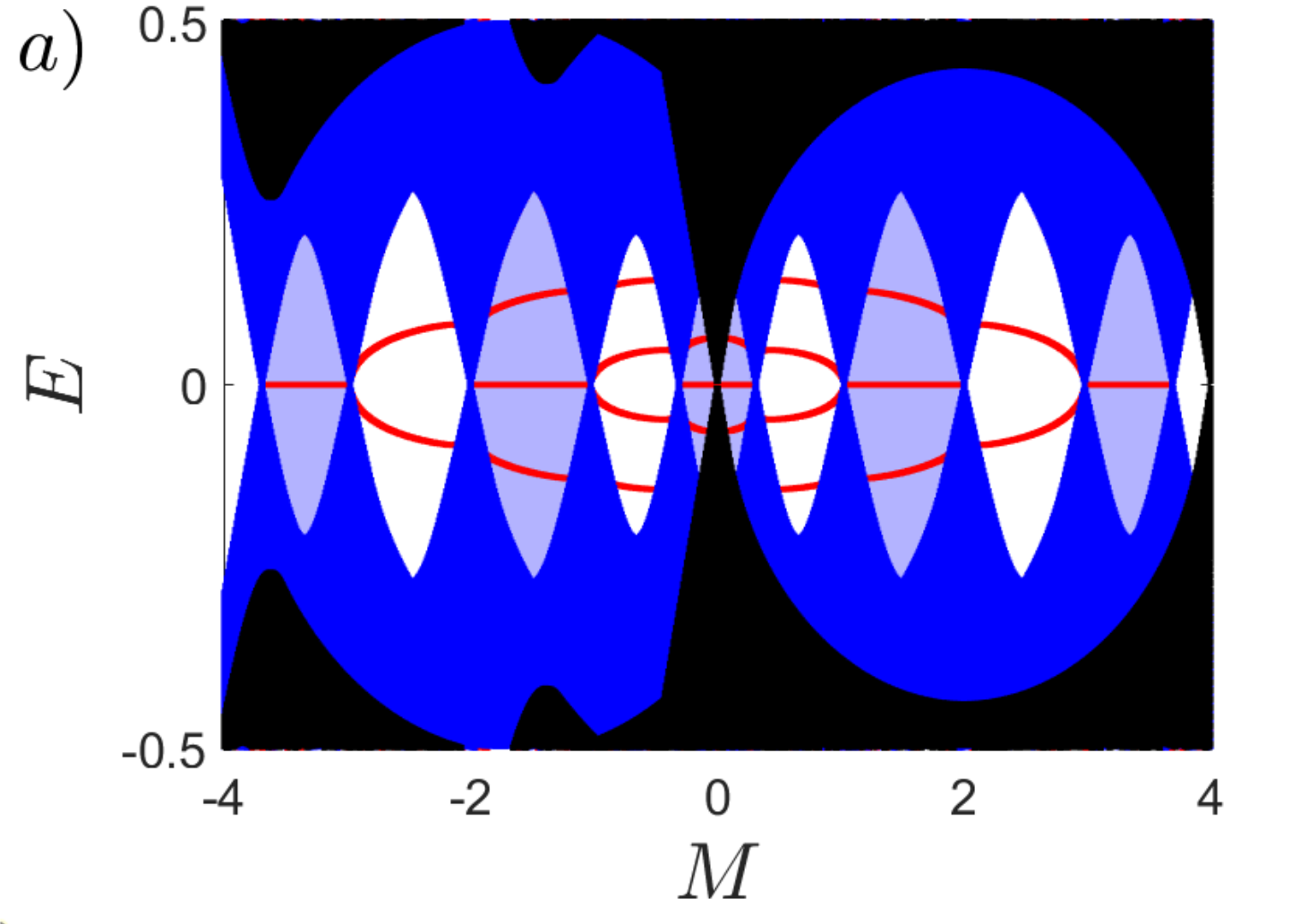}\hfill
\includegraphics[trim=96 240 98 255,clip, width=0.33\linewidth]{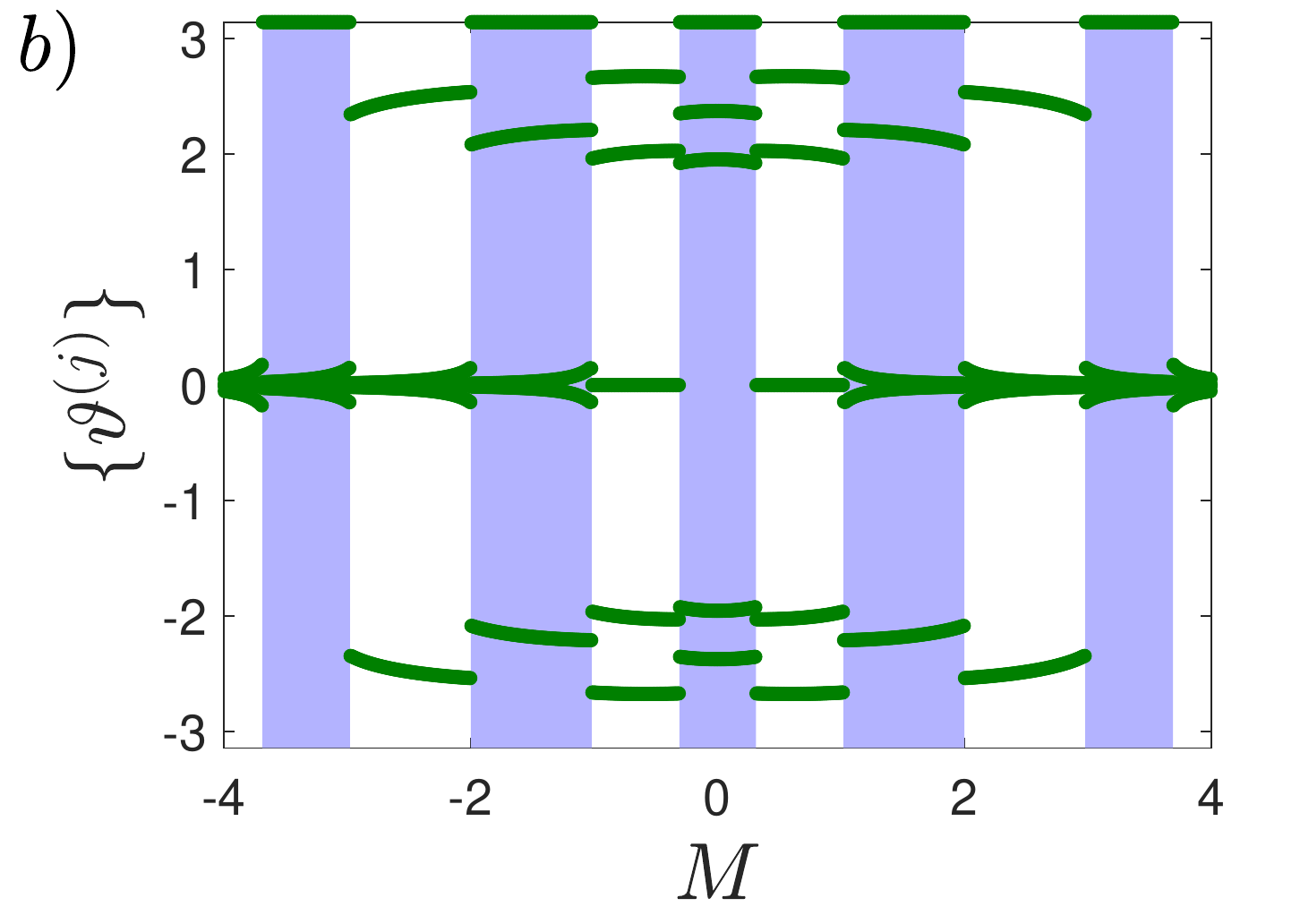}\hfill
\includegraphics[trim=96 240 98 255,clip, width=0.33\linewidth]{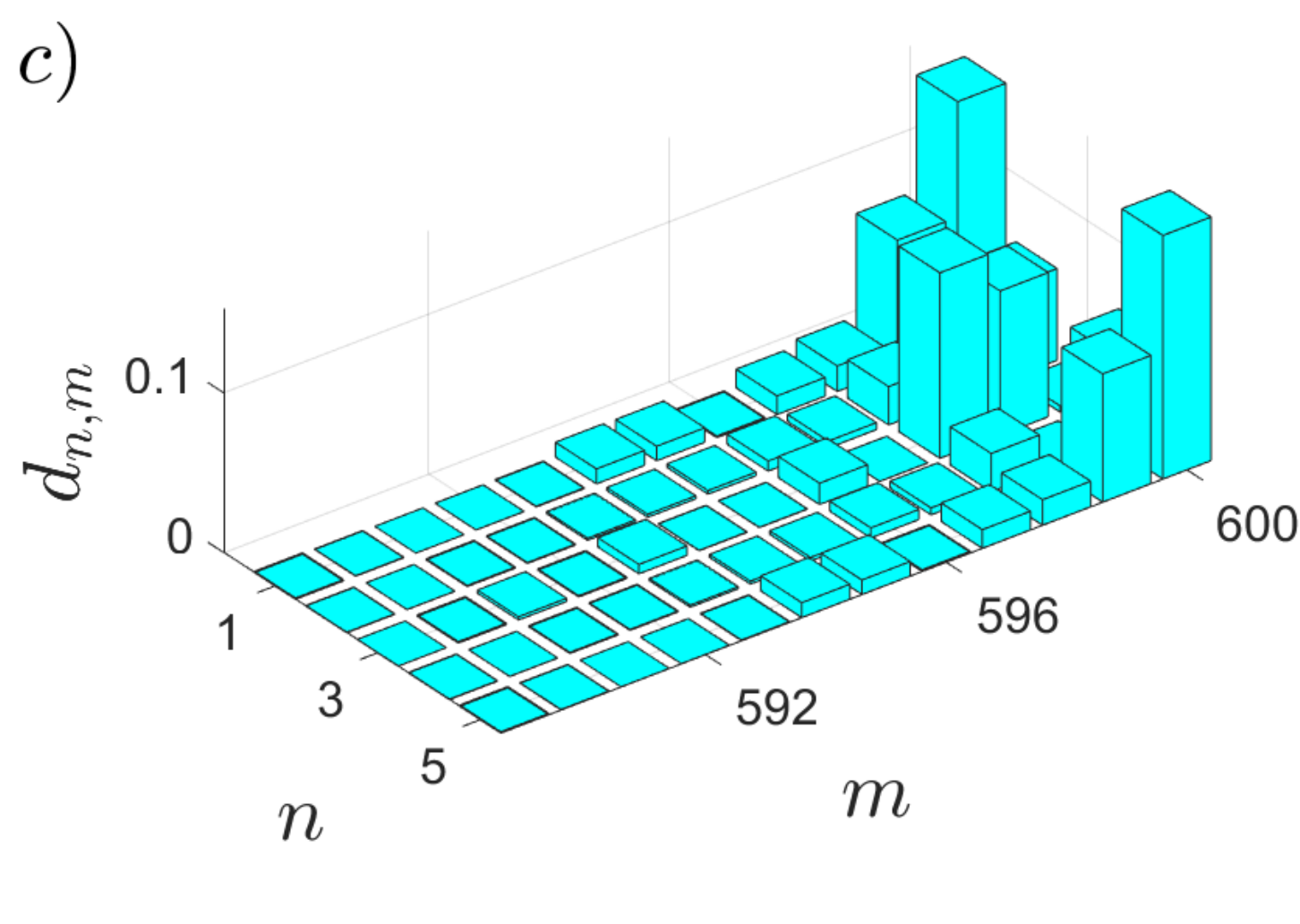}
\caption{(a) Single-particle spectrum of the Chern insulator model in Eq.~(1) in the main text with $L_x=5$, $L_y=600$, $t=1$, $\Delta=0.22$, $\epsilon=0$, and $\kappa_{n,m}=0$. The spectrum for (pe,pe) boundary conditions (black) is plotted on top of the spectrum for (op,pe) boundary conditions (blue), which in turn is plotted on top of the spectrum for (op,op) boundary conditions (red). The purple background highlights the topological regions determined from the Wannier charge centers. (b) Wannier charge centers $\vartheta^{(j)}$ for $L_y\to\infty$. The topological regions (purple) are those for which at least one of the Wannier charge centers are at $\pm \pi$. (c) Density $d_{n,m}=\sum_\sigma\langle c_{\sigma,n,m}^\dag c_{\sigma,n,m} \rangle$ of one of the two states close to zero energy at $M=1.5$. Only one end of the system is shown as more than $99.8\%$ of the total density is located within the sites shown. The other state has the density located at the other end.}
\label{fig:Lx5}
\end{figure*}

The Chern insulator with $L_x=5$ and $L_y=600$ is investigated in Fig.~\ref{fig:Lx5}, where we show the single-particle spectrum, the Wannier charge centers, and the localization of the boundary modes. The results are similar to the case with $L_x=6$, except that the number of bubbles is different. In general, we observe that the (op,pe) spectrum has $2L_x$ gap closings as a function of $M$ when $L_x\ll L_y$, and hence there are $2L_x-1$ bubbles. When counting the bubbles in Fig.~\ref{fig:Lx5}(a), note that the (op,pe) spectrum does not have a gap closing at $M=0$ for $L_x=5$, even though the (pe,pe) spectrum has.

\section{Supplemental Material III. Disorder}\label{Sec:disorder}

The disorder averaging of the Chern insulator spectrum in Fig.~4(b) in the main text is done as follows. We label the energies $E_n$ in the single-particle spectrum with the index $n\in\{1,2,\ldots,2L_xL_y\}$ in increasing order such that $E_{n+1}\geq E_n$. We observe numerically that the states within the bubble gaps form pairs of almost degenerate states with the densities located at opposite ends of the considered quasi-(2-1)D lattice with $6\times 600$ sites. Specifically, we select states that fulfil both of the following criteria and observe that these states coincide with the states in the bubble gaps: Each of the states in a pair with energies $E_n$ and $E_{n+1}$ should have at least $97.5\%$ of the density located within the $2.5\%$ outermost sites at one end of the lattice in more than half of the considered disorder realizations, and for all disorder realizations $\sum_{(n,m)\in\textrm{left}}d_{n,m}-\sum_{(n,m)\in\textrm{right}}d_{n,m}$ should have opposite signs for the two states, where $d_{n,m}=\sum_\sigma \langle  c^\dag_{\sigma,n,m} c_{\sigma,n,m}\rangle$ is the density and left/right refers to the left/right half of the lattice. Depending on the specific disorder realization, one state or the other in a pair may be higher in energy, since the disorder is not the same at the two ends. When we average over different disorder realizations, we want to average a given state with the corresponding states in all the other spectra. Hence, for the states in the bubble gaps, we average the energies of the states that have the densities located at the same end. For all other states, Fig.~1(d) in the main text shows the disorder average of $E_n$.

\begin{figure*}
    \includegraphics[trim=98 240 122 240,clip, width=0.33\textwidth]{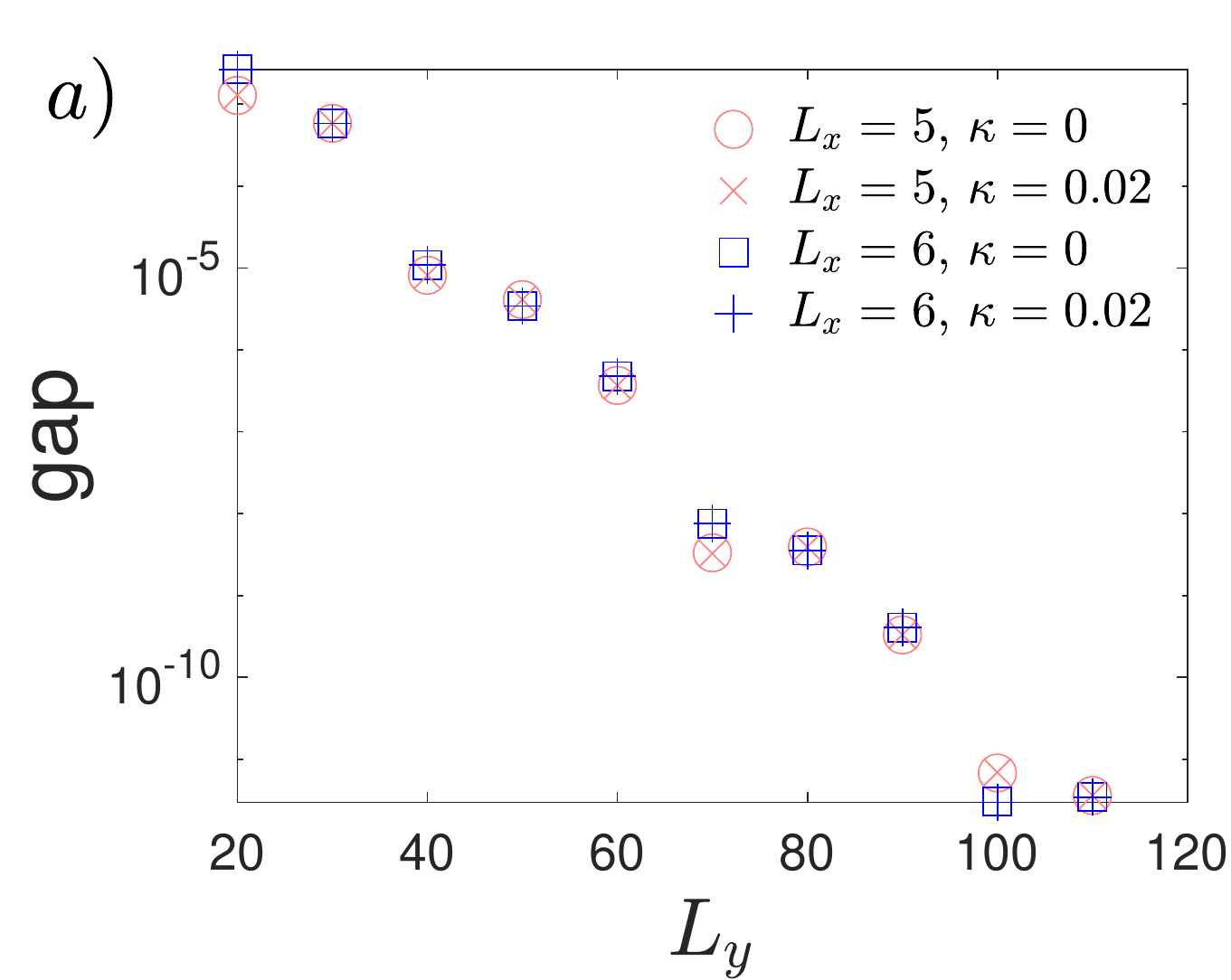}\hfill
    \includegraphics[trim=98 240 122 240,clip, width=0.33\textwidth]{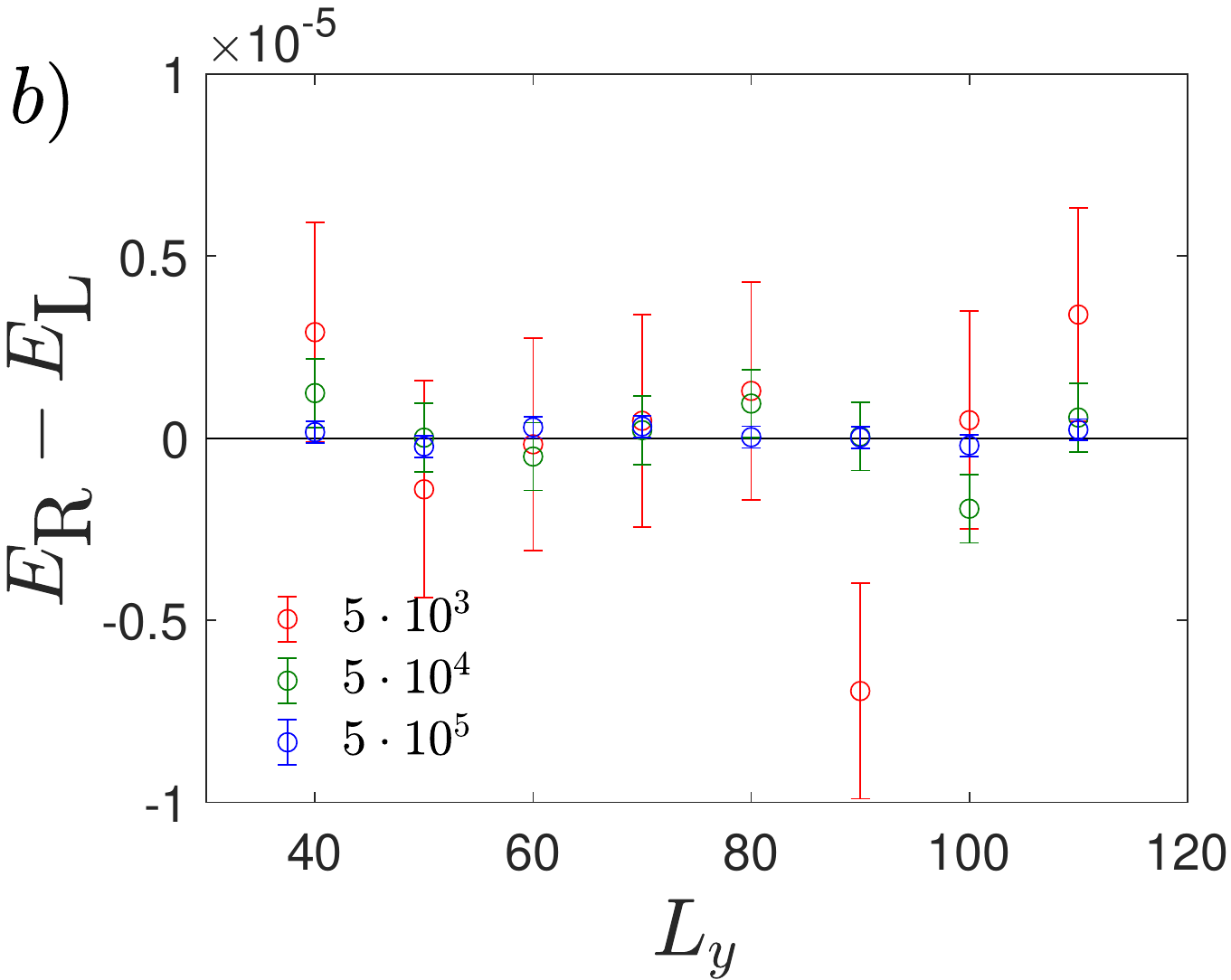}\hfill
    \includegraphics[trim=98 240 122 240,clip, width=0.33\textwidth]{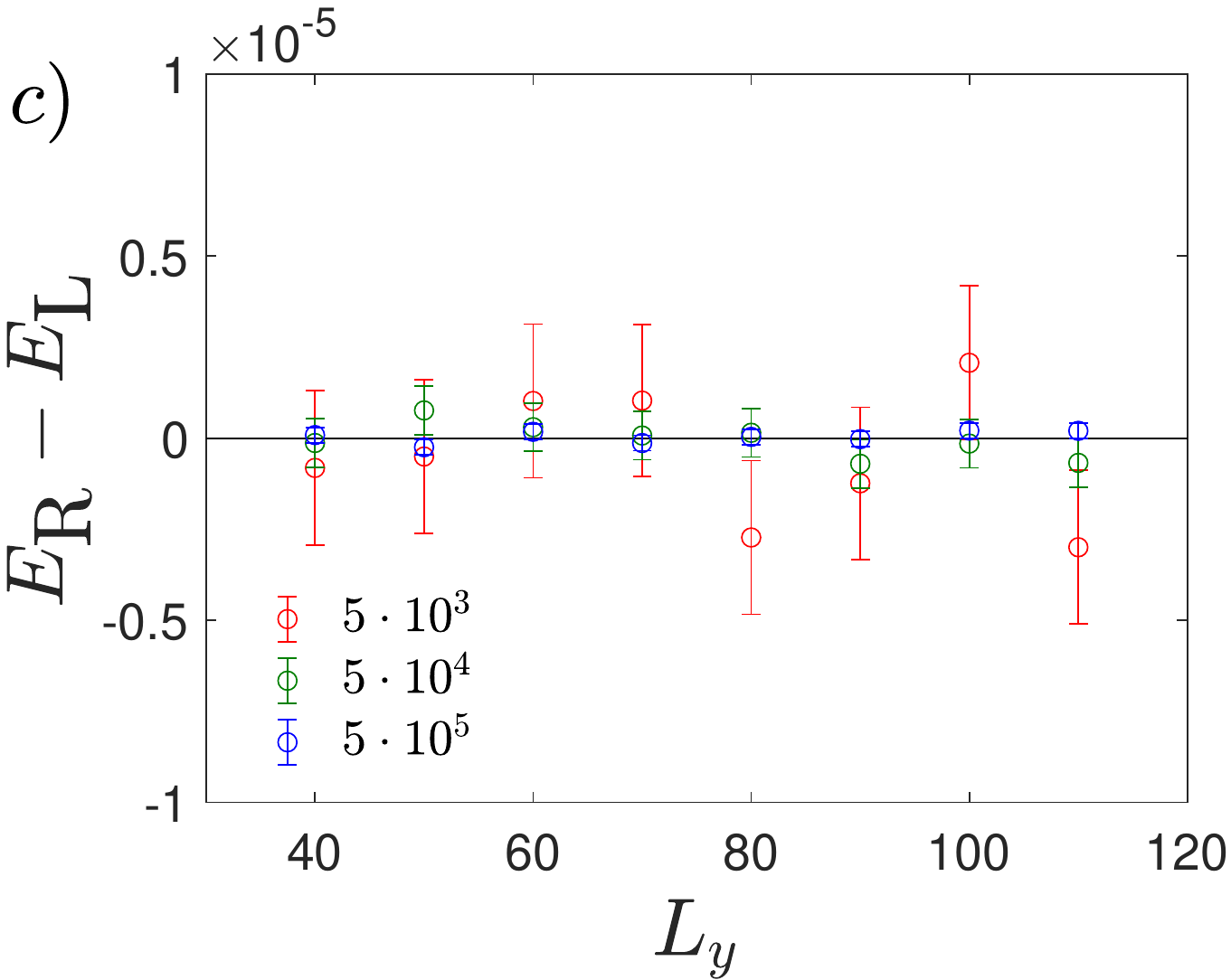}
    \caption{(a) Energy gap between the nearly-degenerate, topological boundary modes (state number $L_xL_y$ and $L_xL_y+1$ in the spectrum) as a function of $L_y$ for the particle-hole symmetric case (i.e.\ $\epsilon=0$) and either $L_x=5$ and $M=1.5$ or $L_x=6$ and $M=2$. The gap for the model with disorder strength $\kappa=0.02$ averaged over $200$ disorder realizations shows the same exponential decay with $L_y$ as the gap for the model without disorder (i.e.\ $\kappa=0$). (b) Gap between the nearly-degenerate, topological boundary modes (state number $L_xL_y$ and $L_xL_y+1$ in the spectrum) for the disordered model without particle-hole symmetry and parameters $L_x=5$, $M=1.5$, $\epsilon=0.02$, and $\kappa=0.02$. The disorder averaging is done as explained in Sec.~\ref{Sec:disorder}, and the gap is shown as the difference between the energy $E_R$ of the state with the density mainly at the right end minus the energy $E_L$ of the state with the density mainly at the left end. The number of disorder realizations is given in the legend, and the error bars show the standard deviation of the mean value. The gap is seen to approach zero as the number of disorder realizations increases. (c) shows the same as (b), but for $L_x=6$ and $M=2$.}
    \label{fig:gapdis}
\end{figure*}

We further investigate the gap between the nearly-degenerate, topological, quasi-(2-2)D boundary modes in the Chern insulator model with disorder in Fig.~\ref{fig:gapdis}. Figure \ref{fig:gapdis}(a) shows the gap for the model with particle-hole symmetry. Even in the presence of disorder, the gap decays exponentially as a function of $L_y$, and after averaging over $200$ disorder realizations, the gaps closely follow the gaps for the model without disorder. The gap for the case without particle-hole symmetry is investigated in Fig.~\ref{fig:gapdis}(b,c). It is seen that the gap approaches zero when the number of disorder realizations increases.

\section{Supplemental Material IV. Quasi-(2-2)D modes in a Chern insulator with Chern number 2}

\begin{figure*}
    \includegraphics[trim=98 240 108 255,clip, width=0.33\textwidth]{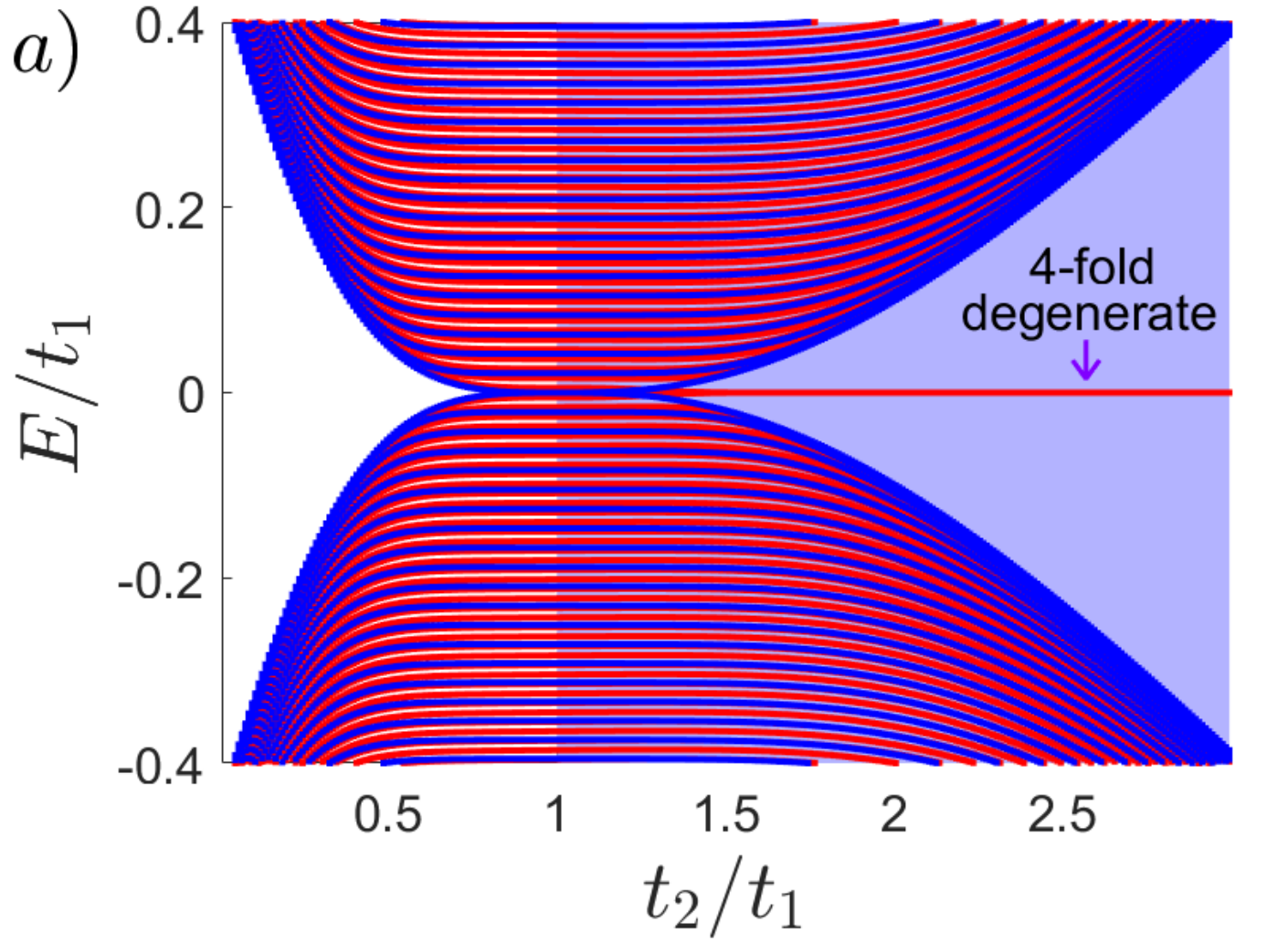}\hfill
    \includegraphics[trim=98 240 108 255,clip, width=0.33\textwidth]{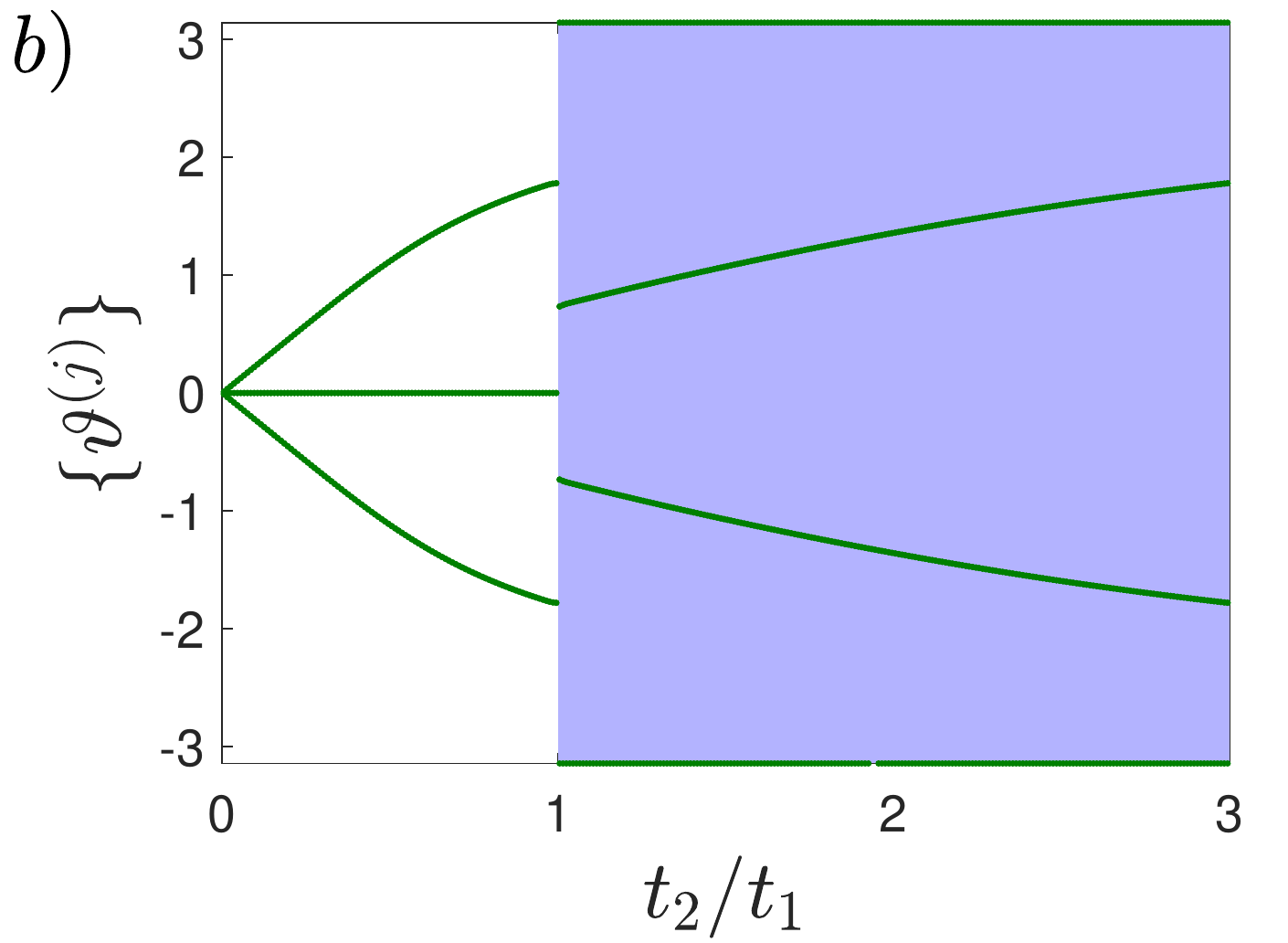}\hfill
    \includegraphics[trim=98 240 108 255,clip, width=0.33\textwidth]{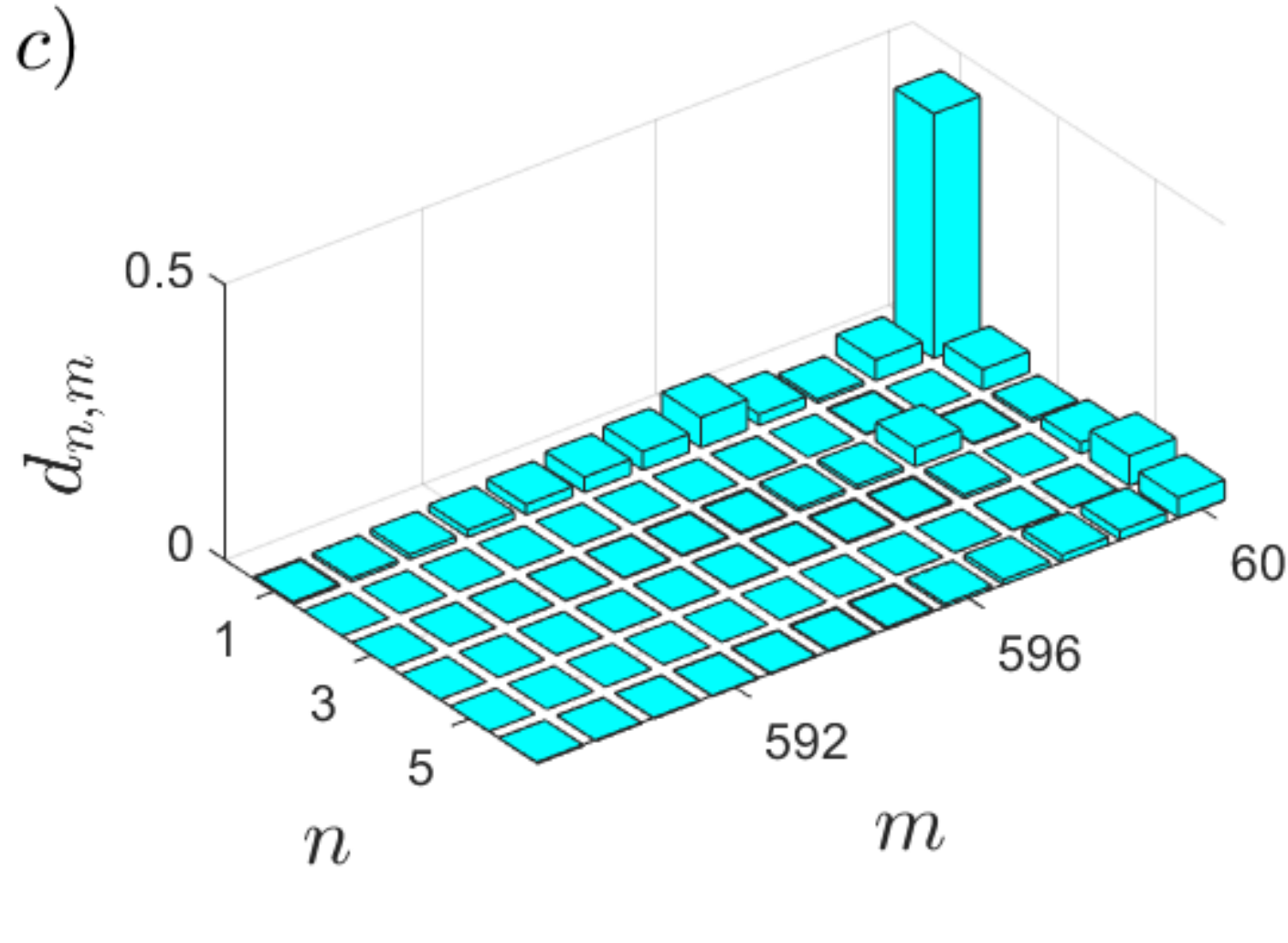}
    \caption{(a) The single-particle spectrum of the model in Eq.~\eqref{Chern2Ham} shows 4-fold degenerate, topological, zero energy modes for $L_x=6$, $L_y=600$, and open boundary conditions in both directions (red). The spectrum for periodic boundary conditions in the $y$ direction and open boundary conditions in the $x$ direction is shown in blue. The purple background highlights the topological region determined from the Wannier charge centers in the limit $L_y\to\infty$. (b) Wannier charge centers $\vartheta^{(j)}$. The topological region (purple) is the region for which at least one of the Wannier charge centers is at $\pm\pi$. Here, we observe that for $t_2>t_1$ and $L_y\to\infty$, two of the Wannier charge centers are at $\pm\pi$. (c) Density $d_{n,m}=\sum_\sigma\langle c^\dag_{\sigma,n,m} c_{\sigma,n,m}\rangle$ near the end of the slab for one of the four zero energy states for $t_2=3t_1$. The total density within the plotted region is $0.988$.}
    \label{fig:C2}
\end{figure*}

In the main text, we considered a Chern insulator model with Chern number $1$ characterized by the Hamiltonian in Eq.~(1). Here, we provide results for a Chern insulator model \cite{Sticlet_2012} on an $L_x\times L_y$ square lattice with two orbitals labeled by $\sigma=\pm1$ on each site and Chern number $2$. The Hamiltonian
\begin{multline}\label{Chern2Ham}
H=t_1 \sum_{\sigma,n,m} (c^\dag_{\sigma,n,m} c_{-\sigma,n+1,m}+
c^\dag_{-\sigma,n+1,m} c_{\sigma,n,m})\\
- t_1 \sum_{\sigma,n,m} \sigma (ic^\dag_{\sigma,n,m} c_{-\sigma,n,m+1}
-ic^\dag_{-\sigma,n,m+1}c_{\sigma,n,m})\\
+ t_2 \sum_{\sigma,n,m} \sigma (c^\dag_{\sigma,n,m} c_{\sigma,n+1,m+1}
+c^\dag_{\sigma,n+1,m+1} c_{\sigma,n,m})
\end{multline}
consists of nearest neighbor hopping between different orbitals with strength $t_1$ and next-nearest neighbor hopping between the same type of orbitals with strength $t_2$. We consider $t_1>0$ and describe the model in terms of the parameter $t_2/t_1$.

We observe different properties of the model with $L_x\ll L_y$ depending on $L_x$. For odd $L_x$, we do not observe interesting behaviors near zero energy. When $L_x$ is divisible by $2$, modes form within a gap in the quasi-(2-1)D system. If $L_x$ is not divisible by $4$, the modes are at zero energy and are 4-fold degenerate. If $L_x$ is divisible by $4$, the modes are at nonzero energy, and there are two sets of states that are each 4-fold degenerate. The degeneracy hence doubles when the Chern number changes from $1$ to $2$. In Fig.\ \ref{fig:C2}, we show results for $L_x=6$. The topological, 4-fold degenerate states at zero energy are seen in the spectrum. The density distributions of these four states reveal the quasi-(2-2)D nature of the states with the densities being mainly at the ends, and the Wannier charge centers show that the model is topological for $t_2>t_1$ in the limit $L_y\to\infty$.

These results suggest that Chern insulators with Chern number $C$ possess $2|C|$, topological, quasi-(2-2)D boundary modes in the topological bubbles. We find this corresponds to $|C|$ centers of the Wannier center spectrum taking the value $\pm\pi$, indicating integer topological classification for the finite-size system.

\bibliography{p1bib.bib}

\end{document}